\documentclass[twocolumn]{aastex631}
\usepackage{natbib}
\usepackage{amsmath}
\usepackage{verbatim}
\usepackage{hyperref}
\usepackage{float}
\usepackage{graphicx}
\usepackage{mathtools}
\usepackage{empheq}
\usepackage{appendix}
\usepackage{microtype}
\usepackage{commath}
\usepackage{bold-extra}
\usepackage{comment}
\usepackage{rotating}
\DeclareGraphicsExtensions{.pdf,.png,.jpg}
\DeclareUnicodeCharacter{2223}{$|$}

\newcommand{\bq}{\begin{equation}} 
\newcommand{\eq}{\end{equation}} 
 
\newcommand{\hunit}{km\,s$^{-1}$\,Mpc$^{-1} \, $}

\defcitealias{MF2025}{MF25}
\defcitealias{Lindegren2021_plx_bias}{L21}
\defcitealias{Benedict2007}{B07}

\definecolor{myblue}{RGB}{0, 170, 225} 
\definecolor{myred}{RGB}{215, 0, 0}

\shorttitle{Breuval et al.}
\shortauthors{Breuval et al. }

\begin{document}


\title{Converging on the Cepheid Metallicity Dependence: \\ Implications of Non-Standard Gaia Parallax Recalibration on Distance Measures}



\author[0000-0003-3889-7709]{Louise Breuval}
\altaffiliation{ESA Research Fellow}
\affiliation{European Space Agency (ESA), ESA Office, Space Telescope Science Institute, 3700 San Martin Drive, Baltimore, MD 21218, USA}
\email{lbreuval@stsci.edu}

\author[0000-0002-5259-2314]{Gagandeep S. Anand}
\affiliation{Space Telescope Science Institute, 3700 San Martin Drive, Baltimore, MD 21218, USA}

\author[0000-0001-8089-4419]{Richard I.~Anderson}
\affiliation{Institute of Physics, \'Ecole Polytechnique F\'ed\'erale de Lausanne (EPFL), Observatoire de Sauverny, 1290 Versoix, Switzerland}

\author[0000-0002-1691-8217]{Rachael Beaton}
\affiliation{Space Telescope Science Institute, 3700 San Martin Drive, Baltimore, MD 21218, USA}

\author[0000-0001-6147-3360]{Anupam Bhardwaj}
\affiliation{Inter-University Centre for Astronomy and Astrophysics (IUCAA), Post Bag 4, Ganeshkhind, Pune 411 007, India}

\author[0000-0000-0000-0000]{Stefano Casertano}
\affiliation{Space Telescope Science Institute, 3700 San Martin Drive, Baltimore, MD 21218, USA}

\author[0000-0001-9206-9723]{Gisella Clementini}
\affiliation{INAF, Osservatorio di Astrofisica e Scienza dello Spazio di Bologna, Via Piero Gobetti 93/3, 40129 Bologna, Italy}

\author[0000-0003-2443-173X]{Mauricio Cruz Reyes}
\affiliation{Institute of Physics, \'Ecole Polytechnique F\'ed\'erale de Lausanne (EPFL), Observatoire de Sauverny, 1290 Versoix, Switzerland}

\author[0000-0002-5819-3461]{Giulia De Somma}
\affiliation{INAF – Osservatorio Astronomico di Capodimonte, Via Moiariello 16, 80131 Napoli, Italy}
\affiliation{INAF – Osservatorio Astronomico d'Abruzzo, Via Maggini sn, 64100 Teramo, Italy}
\affiliation{Istituto Nazionale di Fisica Nucleare (INFN), Sezione di Napoli, Compl. Univ. di Monte S. Angelo,  \\ Edificio G, Via Cinthia, I-80126 Napoli, Italy}

\author[0000-0003-2723-6075]{Martin A. T. Groenewegen}
\affiliation{Koninklijke Sterrenwacht van Belgi\"e, Ringlaan 3, B-1180 Brussels, Belgium}

\author[0000-0001-6169-8586]{Caroline D. Huang}
\altaffiliation{NSF Astronomy and Astrophysics Postdoctoral Fellow}
\affiliation{Center for Astrophysics $\vert$ Harvard \& Smithsonian, 60 Garden Street, Cambridge, MA 02138, USA}

\author[0000-0003-0626-1749]{Pierre Kervella}
\affiliation{LIRA, Observatoire de Paris, Universit\'e PSL, Sorbonne Universit\'e, Universit\'e Paris Cit\'e, \\ CY Cergy Paris Universit\'e, CNRS, 5 place Jules Janssen, 92195 Meudon, France}
\affiliation{French-Chilean Laboratory for Astronomy, IRL 3386, CNRS and U. de Chile, Casilla 36-D, Santiago, Chile}

\author[0000-0001-5998-5885]{Saniya Khan}
\affiliation{Institute of Physics, \'Ecole Polytechnique F\'ed\'erale de Lausanne (EPFL), Observatoire de Sauverny, 1290 Versoix, Switzerland}

\author[0000-0002-1775-4859]{Lucas M.~Macri}
\affiliation{Department of Physics \& Astronomy, College of Sciences, University of Texas Rio Grande Valley,\\ 1201 W University Dr, Edinburg TX 78539, USA}

\author[0000-0002-1330-2927]{Marcella Marconi}
\affiliation{INAF – Osservatorio Astronomico di Capodimonte, Via Moiariello 16, 80131 Napoli, Italy}

\author[0000-0002-6342-0331]{Javier H. Minniti}
\affiliation{Department of Physics and Astronomy, Johns Hopkins University, Baltimore, MD 21218, USA}

\author[0000-0002-6124-1196]{Adam G.~Riess}
\affiliation{Space Telescope Science Institute, 3700 San Martin Drive, Baltimore, MD 21218, USA}
\affiliation{Department of Physics and Astronomy, Johns Hopkins University, Baltimore, MD 21218, USA}

\author[0000-0003-1801-426X]{Vincenzo Ripepi}
\affiliation{INAF – Osservatorio Astronomico di Capodimonte, Via Moiariello 16, 80131 Napoli, Italy}

\author[0000-0002-5527-6317]{Martino Romaniello}
\affiliation{European Southern Observatory, Karl-Schwarzschild-Strasse 2, 85478 Garching bei München, Germany}

\author[0000-0002-4934-5849]{Daniel Scolnic}
\affiliation{Department of Physics, Duke University, Durham, NC 27708, USA}

\author[0000-0002-3899-566X]{Erasmo Trentin}
\affiliation{INAF – Osservatorio Astronomico di Capodimonte, Via Moiariello 16, 80131 Napoli, Italy}

\author[0000-0002-1662-5756]{Piotr Wielg\'orski}
\affiliation{Nicolaus Copernicus Astronomical Center, Polish Academy of Sciences, Bartycka 18, 00-716 Warszawa, Poland}

\author[0000-0001-9420-6525]{Wenlong Yuan}
\affiliation{Department of Physics and Astronomy, Johns Hopkins University, Baltimore, MD 21218, USA}

\begin{abstract}
By comparing Cepheid brightnesses with geometric distance measures including {\it Gaia} EDR3 parallaxes, most recent analyses conclude metal-rich Cepheids are brighter, quantified as $\gamma \sim -0.2$ mag/dex. While the value of $\gamma$ has little impact on the determination of the Hubble constant in contemporary distance ladders (due to the similarity of metallicity across these ladders), $\gamma$ plays a role in gauging the distances to metal-poor dwarf galaxies like the Magellanic Clouds and is of considerable interest in testing stellar models. Recently, \citet[hereafter MF25]{MF2025} recalibrated \textit{Gaia} EDR3 parallaxes by adding to them a magnitude offset to match certain historic Cepheid parallaxes which otherwise differ by $\sim  1.6\sigma$. A calibration which adjusts \textit{Gaia} parallaxes by applying a magnitude offset (i.e., a multiplicative correction in parallax) differs significantly from the \textit{Gaia} Team’s calibration \citep{Lindegren2021_plx_bias}, which is additive in parallax space — especially at distances much closer than 1 kpc or beyond 10 kpc, outside the $\sim$2–3 kpc range on which the \citetalias{MF2025} calibration was based. The \citetalias{MF2025} approach reduces $\gamma$ to zero. If broadly applied, it places nearby cluster distances like the Pleiades too close compared to independent measurements, while leaving distant quasars with negative parallaxes. We conclude that the \citetalias{MF2025} proposal for \textit{Gaia} calibration and $\gamma \sim 0$ produces farther-reaching consequences, many of which are strongly disfavored by the data. 
\end{abstract}

\section{Introduction}
\label{sec:intro}

\begin{figure}[h!]
\centering
\includegraphics[height=13.0cm]{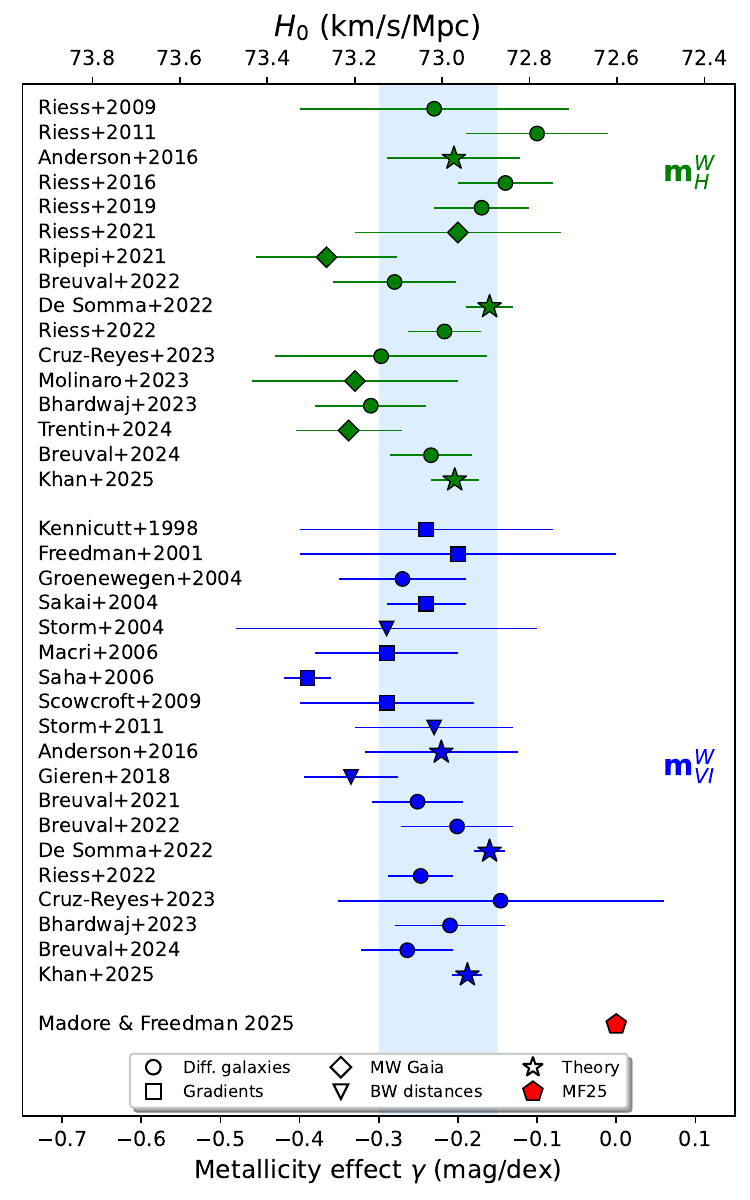}  
\caption{Recent empirical and theoretical estimates of the Cepheid metallicity dependence $\gamma$ from the literature in the $m_H^W$ and $m_{VI}^W$ Wesenheit indices. A few analyses were excluded from this plot \citep[e.g.][]{Freedman2011, Wielgorski2017, Owens2022} due to issues that were identified and discussed in  \citet{Breuval2022}. The blue band shows broad agreement between $\gamma \sim -0.15$ and $-0.30$ mag/dex. The methods labeled with different shapes include: comparison between P--L relations in different galaxies, metallicity gradients, Milky Way Cepheids with {\it Gaia} EDR3 parallaxes, Baade-Wesselink distances, and theoretical predictions. The red point indicates the $\gamma \sim 0$ multi-wavelength result by \citetalias{MF2025}, based on a variety of methods. The metallicity dependence $\gamma$ in the $m_H^W$ Wesenheit magnitude and the Hubble constant $H_0$ \citep[derived from three anchor galaxies,][]{Riess2022a} are related as follows: $\gamma$ shifts the luminosity of Cepheids which show the largest metallicity difference with respect to Cepheids in SNIa hosts (i.e. in the SH0ES distance ladder, the metal-poor LMC and SMC). This luminosity difference directly translates into a shift in $H_0$. We note that the $H_0$ values represented on the top $x$-axis are
not actually measured in the quoted references but are derived from \citet{H0DN2025}. \\ }   
\label{fig:gamma_consensus}
\end{figure}

The dependence of the Cepheid Period--Luminosity (P--L) relation on metallicity is astrophysical in nature, quantified as $\gamma$ (in mag/dex) and is crucial when using Cepheid brightnesses to determine their distances. Since the time of the final result of the HST Key Project \citep{Freedman2001,Kennicutt1998}, this term was found to be negative, indicating that at fixed period and color, metal-rich Cepheids are intrinsically brighter. \cite{Freedman2001} write: {\it ``Published empirical values for the index $\gamma$ range from 0 to $-1.3$ mag/dex (with most values between 0 and $-0.4$) [...] Other recent studies conclude that a metallicity effect is extant, and all of the empirical studies agree on the sign, if not the magnitude of the effect. Considering all of the evidence currently available and the (still considerable) uncertainties, we therefore adopt $\gamma = -0.2 \pm 0.2 \, \rm mag/dex$, approximately the midrange of current empirical values, and correct our Cepheid distances accordingly''}. Although its exact value is still debated, the community has now reached a broad, though not unanimous, consensus on the negative sign and on the value of this effect: $\gamma \sim -0.2$ mag/dex with variations of about $\pm$0.1 mag/dex depending on the study (see Fig.~\ref{fig:gamma_consensus}). 

A metallicity dependence of Cepheid pulsation properties is theoretically expected. Stellar evolution models predict that, at fixed mass, metal-rich Cepheids are slightly fainter than their metal-poor counterparts \citep{CastellaniChieffiStraniero1992, BonoCaputoCassisi2000}, with helium abundance showing the opposite effect. In contrast, stellar pulsation models \citep{BonoCastellaniMarconi2000, Marconi2005, Anderson2016, DeSomma2022, Khan2025} predict that an increase in metallicity shifts the instability strip to redder colors at constant stellar mass, as the enhanced opacity reduces pulsation driving in the hydrogen ionization zone \citep[see][for details]{BonoMarconi1999}. Conversely, a higher helium abundance shifts the instability strip in the opposite direction \citep{Fiorentino2002, Marconi2005}. These dependencies directly affect the P--L and Period-Wesenheit relations, and all recent models consistently predict that, at fixed period and color, metal-rich Cepheids are intrinsically brighter than metal-poor ones ($\gamma < 0$). We further note that the modelled atmosphere results presented in \citet{Madore2025_theory} do not change the above scenario, because a metallicity dependence is already expected in bolometric magnitudes as an effect of the pulsation mechanism physics, and is not only an atmospheric phenomenon.

In order to calibrate Cepheid distances in the Milky Way, to estimate the metallicity dependence of the P--L relation, to measure extragalactic distances and ultimately the local value of the Hubble Constant ($H_0$), many recent empirical studies \citep[e.g.][]{ZhouChen2021, Riess2021, Breuval2021, Breuval2022, Riess2022b, Owens2022, Ripepi2022, Ripepi2022Gaia, Bhardwaj2023, Bhardwaj2024, CruzReyes2023, Trentin2024} rely on {\it Gaia} Early Data Release 3 (EDR3), which provides stellar parallaxes for over 1.5 billion sources with unprecedented accuracy and completeness \citep{GaiaCollaboration2021}. \textit{Gaia} EDR3 parallaxes require additional calibration at the level of $\sim20 \, \mu$as.  A primary source of calibration comes from \textit{Gaia} EDR3 astrometry of quasars, which reveals parallaxes that are negative by a few tens of microarcseconds. Additional calibration comes from stars in the Large Magellanic Cloud (LMC), with extension to brighter magnitudes from physical pairs. Based on an extensive, simultaneous analysis of this data, \citet[hereafter L21]{Lindegren2021_plx_bias} provides a calibration or parallax offset correction (hereafter $\varpi_{\rm L21}$), an angular term which is a function of magnitude, color and ecliptic latitude. This parallax correction -- which produces the most accurate results -- involves subtracting\footnote{\citetalias{Lindegren2021_plx_bias}: ``{\it Regarded as a systematic correction to the parallax, the bias function Z5 or Z6 should be subtracted from the value (parallax) given in the archive. Python implementations of both functions are available in the Gaia web pages}”.} a value in angular units derived by the provided function\footnote{\href{https://www.cosmos.esa.int/web/gaia/edr3-code}{https://www.cosmos.esa.int/web/gaia/edr3-code}}, and has been corroborated by independent studies \citep[e.g.][]{Groenewegen2021, MaizApellaniz2021}. Various works have investigated the need for an additional offset in the brighter range of magnitudes ($G < 10$ mag) where the \citetalias{Lindegren2021_plx_bias} analysis has limited sampling \citep{Bhardwaj2021, Fabricius2021, Huang2021, Ren2021, Stassun2021, Vasiliev2021, Zinn2021, Khan2023}. For a sample of 75 Milky Way Cepheids ($G<8$ mag), \cite{Riess2021} find a best-fit residual offset (hereafter $\varpi_{\rm offset}$) after application of the \citetalias{Lindegren2021_plx_bias} correction, of $\varpi_{\rm offset} = -14 \pm 6 \, \rm \mu as$, in the sense that \citetalias{Lindegren2021_plx_bias} slightly overcorrects parallaxes. This step does not depend on any external parallax reference, and is based only on minimizing the P--L dispersion. Other independent studies have obtained similar values \citep[][with $\varpi_{\rm offset} = -19 \pm 3 \, \rm \mu$as, $\varpi_{\rm offset} = -15 \pm 3 \, \rm \mu$as, and $\varpi_{\rm offset} < 0 \, \mu$as, respectively]{CruzReyes2023, Wang2024, Trentin2024}. Overall, adopting the \citetalias{Lindegren2021_plx_bias} parallax calibration $\varpi_{\rm L21}$ as well as a $\varpi_{\rm offset} = -14\, \mu$as counter-correction for Cepheids results in a metallicity dependence $\gamma$ consistent with $-0.2 \pm 0.1$ mag/dex.

\begin{table*}[t!]
\centering
\caption{Key difference in the treatment of {\it Gaia} EDR3 parallaxes between the recommended calibration by the {\it Gaia} Team \citepalias{Lindegren2021_plx_bias} and the \citetalias{MF2025} work. \\  }
\begin{tabular}{c | c}
\hline
\hline
{\it Gaia} Collaboration \citepalias{Lindegren2021_plx_bias} & \citet{MF2025} \\
\hline
Correction in Parallax Space & Correction in Magnitude Space \\
                             & (m-M)$_{\rm corr}$ = (m-M)$_{\rm Gaia \, EDR3} -0.26$ \\
\hline
{\bf Additive} Parallax Offset     & Equiv. to {\bf Multiplicative} Parallax Offset \\
$\varpi_{\rm corr} = \varpi_{\rm Gaia \, EDR3}$ + L21 & $\varpi_{\rm corr} = 1.127 \times \varpi_{\rm Gaia \, EDR3}$ \\
\hline
\multicolumn{2}{c}{These two different corrections have consequences on distance to Pleiades,} \\
\multicolumn{2}{c}{quasar parallaxes, and the inferred measurement of $H_0$} \\
\hline
\multicolumn{2}{c}{} \\
\end{tabular}
\label{table:summary}
\end{table*}

Instead of individual Cepheids, Milky Way open clusters have been used to achieve better astrometric precision for the Cepheids they host, because their parallaxes can be averaged over a large number of stars \citep{Anderson2013, Breuval2020, Riess2022b, CruzReyes2023}. However, since cluster members are located in the same region of the sky, their parallaxes are highly correlated. As a result, even though the statistical uncertainty of a mean cluster parallax can be as low as $2 \, \mu$as, the cluster parallax precision is dominated by the angular covariance, currently estimated to $\sim7 \, \mu$as \citep{MaizApellaniz2021, Vasiliev2021}. The other advantage of using cluster Cepheids is that cluster members are generally fainter ($G>13$ mag), and thus have good overlap with the magnitude and color range of the \citetalias{Lindegren2021_plx_bias} calibration, reducing the size or need for residual parallax offsets. \cite{Riess2022b} and \cite{CruzReyes2023} used cluster Cepheids and found no evidence of residual offset beyond the \citetalias{Lindegren2021_plx_bias} correction in this magnitude range. Similarly, \cite{Wang2024} adopt the largest sample of Milky Way cluster Cepheids to date along with their mean \textit{Gaia} EDR3 parallaxes and find a residual parallax offset consistent with zero for cluster members. In conclusion, using cluster Cepheids with only the \citetalias{Lindegren2021_plx_bias} correction returns similar results compared with using field Cepheids with \citetalias{Lindegren2021_plx_bias} and the additional $-14 \, \mu$as: both yield a metallicity dependence $\gamma$ consistent with $-0.2$ mag/dex, and a distance modulus to the LMC in excellent agreement with the geometric distance by \cite{Pietrzynski2019}.

\citet[hereafter MF25]{MF2025} offer a very different approach to the preceding analyses (see Table~\ref{table:summary}). In \citetalias{MF2025}, the earlier generation of parallax measurements from the Hubble Space Telescope (HST) Fine Guidance Sensor (FGS) from \citet[hereafter B07]{Benedict2007} is used to recalibrate those from \textit{Gaia} EDR3 and to conclude that metallicity has no statistically significant effect, $\gamma \sim 0$, on the Cepheid P--L (and Period-Wesenheit) relation across various wavelengths. 
While the \citetalias{MF2025} conclusion, $\gamma\sim0$, is drawn from a broad dataset and multiple tests, several aspects warrant a critical examination to assess the robustness of this claim. In particular, instead of applying the additive \citetalias{Lindegren2021_plx_bias} calibration in parallax space, \citetalias{MF2025} subtract a fixed magnitude of 0.26 mag from each Cepheid distance modulus to produce a specific P--L scatter when their {\it Gaia} sample is combined with older parallax measures of 10 Cepheids from \citetalias{Benedict2007}. An offset in magnitude space acts as a multiplicative correction in parallax, which based on the way \textit{Gaia} parallaxes are measured (closed phase, and imperfect basic angle monitor due to thermal variations and geometric distortions producing small astrometric shifts) would not be the expected mathematical operation to calibrate any \textit{Gaia} parallaxes \citep[empirical studies of the residual parallax bias as a function of apparent magnitude, e.g.,][and references therein, rule out a simple multiplicative model]{Khan2023}. The \citetalias{MF2025} calibration is applied to the distance modulus:
\begin{equation}
(m-M)_{\rm MF25} = (m-M)_0 - 0.26
\label{eq_MF_dm}
\end{equation}
where the initial distance modulus is defined as $(m-M)_0 = 10 - 5 \log (\varpi_{\rm Gaia \, EDR3})$ with $\varpi_{\rm Gaia \, EDR3}$ the {\it Gaia} EDR3 parallax. As a consequence, the \citetalias{MF2025} approach is equivalent to using a parallax $\varpi_{\rm MF25}$, where:
\begin{equation}
\varpi_{\rm MF25} = 1.127 \, \, \varpi_{\rm Gaia \, EDR3}
\label{eq_MF_plx}
\end{equation}
On the other hand, the parallax calibration recommended by the {\it Gaia} team is:
\begin{equation}
\varpi_{\rm Gaia, \, corr} = \varpi_{\rm Gaia \, EDR3} + \rm \varpi_{\rm L21} 
\label{eq_gaia_plx}
\end{equation}
where $\varpi_{\rm L21}$ is the \citetalias{Lindegren2021_plx_bias} correction. These two approaches are summarized in Table~\ref{table:summary}.


In Sect. \ref{sec:MF_findings}, we inspect the multiple tests carried out by \citetalias{MF2025} and, where possible, point out more up-to-date datasets available for the same analyses, ensuring consistency with recent developments in the field.  While a multiplicative and additive calibration may nearly coincide at some parallax, they will be unequal over a wider range.  Therefore, in Sect. \ref{sec:consequences}, we discuss how the \citetalias{MF2025}  \textit{Gaia} calibration would perform nearby and far away, as well as for the local value of the Hubble Constant. We conclude with a discussion in Sect. \ref{sec:discussion}.  \\

\section{Revisiting the \citetalias{MF2025} Finding of a Null Cepheid Metallicity Dependence}
\label{sec:MF_findings}

A departure from most studies in the last $\sim$ 25 years to claim that the metallicity dependence is consistent with zero, as in \citetalias{MF2025}, requires strong evidence and identifying why nearly all recent studies (Fig.~\ref{fig:gamma_consensus}) find a systematically negative dependence. In this section we investigate the methods employed in \citetalias{MF2025} that lead to the conclusion of a null metallicity dependence. \\

\subsection{The \citetalias{MF2025} recalibration of Gaia EDR3 parallaxes}

The key to the difference in conclusion reached in \citetalias{MF2025} concerning $\gamma$ lies in the unusual treatment of \textit{Gaia} EDR3 parallaxes in that work. Most contemporary studies treat \textit{Gaia} EDR3 as a generational improvement in parallax precision and scope, warranting replacement of all prior measurements. After applying the recommended \citetalias{Lindegren2021_plx_bias} calibration function which is additive in parallax and of size $\sim 20 \, \mu \rm as$, the systematic accuracy in the well-calibrated region is estimated at ``a few micro-arcseconds''. As evidence, \citetalias{Lindegren2021_plx_bias} show a measure of the LMC parallax consistent with the expectation from detached eclipsing binaries \citep[hereafter DEBs,][]{Pietrzynski2019} to within a few micro-arcseconds and the AGN catalogue has a mean parallax of 0.5 $\mu \rm as$.

In contrast, \citetalias{MF2025} identify an offset between 1) the Cepheid P--L relation for \textit{Gaia} EDR3 parallaxes of Cepheids in clusters {\it before application of the \citetalias{Lindegren2021_plx_bias} calibration term}, and 2) the Cepheid P--L relation for 10 nearby Cepheids with previous generation HST FGS parallaxes from \citetalias{Benedict2007}. \citetalias{MF2025} discard the \citetalias{Lindegren2021_plx_bias} parallax calibration and apply a magnitude offset of 0.26 mag to each Cepheid distance modulus, so that the P--L dispersion of the cluster Cepheid sample and HST FGS sample combined matches the P--L dispersion obtained in the LMC\footnote{ \citetalias{MF2025}: ``{\it the data are fit to an a priori dispersion, rather than to a minimum dispersion. [...] We determine this offset only once (at 3.6 $\mu$m), and next apply it equally to all other bands, without modification.}''}. The \citetalias{Benedict2007} parallaxes are an order of magnitude less precise ($200-300 \, \mu \rm as$) than those from {\it Gaia} EDR3, allowing meaningful parallax measurements for only the nearest $\sim$10 Cepheids, but also a factor of 10 less precise than {\it Gaia}'s residual parallax bias, which disqualifies them from providing strong constraints. Additionally, the magnitude range of Cepheids in \citetalias{Benedict2007} is too bright for good {\it Gaia} parallaxes, precluding an immediate comparison in parallax space, hence \citetalias{MF2025} used P--L relations instead. Doing the comparison in magnitude/P--L relation space, as in \citetalias{MF2025}, brings additional complications and leads to the interpretation of {\it Gaia}'s systematics as a multiplicative term, which would have wide-ranging consequences that were not considered in \citetalias{MF2025} and are addressed here. Finally, we note that \citetalias{Benedict2007} HST FGS parallaxes are infrequently used in current P--L calibrations, as they have been superseded by \textit{Gaia} EDR3 parallaxes. For example, \citet{Gallenne2025} show that the P--L relation based on \citetalias{Benedict2007} parallaxes yields inaccurate distances compared to precise geometric measurements. Therefore, it might seem questionnable to require the astrometrically more accurate and widely tested \textit{Gaia} EDR3 parallaxes to match the less accurate.

Adopting the \citetalias{MF2025} correction rather than the \citetalias{Lindegren2021_plx_bias} calibration assumes three elements: 1) that the 1.6$\sigma$ difference between the \textit{Gaia} EDR3 and the \citetalias{Benedict2007} samples is significant and warrants a parallax recalibration, 2) that any parallax inaccuracy lies with \textit{Gaia} EDR3 rather than the 10 HST FGS parallaxes from \citetalias{Benedict2007}, and 3) that matching the LMC geometric distance without a Cepheid metallicity term is preferable. Thus, bringing \textit{Gaia} EDR3 parallaxes of clusters hosting Cepheids into better agreement with the HST FGS parallax sample reduces the metallicity term $\gamma$ to zero, but with what consequences (see Sect.~\ref{sec:consequences})?  \\

\begin{table*}[t!]
\small
\centering
\caption{Successive updates of the Milky Way cluster Cepheid P--L relation, $M_{[3.6]} = \alpha \, (\log P-1) + \beta$, from \citetalias{MF2025}. Columns 1 and 2 give the P--L intercept and scatter respectively, column 3 gives the inferred metallicity dependence $\gamma$ and the last column describes the improvements corresponding to each row.  \\}
\begin{tabular}{c c c l}
\hline
\hline
$\beta$ & $\sigma$ & $\gamma$ & Comments  \\
 (mag)   & (mag)    & (mag/dex)    &  \\
\hline
$-5.823 \pm 0.022$ & 0.071 & $-0.06 \pm 0.05$ & Initial \citetalias{MF2025} calibration including 0.26 mag offset  \\  
$-5.865 \pm 0.024$ & 0.078 & $-0.14 \pm 0.05$ & Replace $(m-M)_{0, \, \rm MF25}$ with \textit{Gaia} EDR3 parallaxes + \citetalias{Lindegren2021_plx_bias} \\   
$-5.874 \pm 0.023$ & 0.080 & $-0.16 \pm 0.05$ & Add V367 Sct with correct period \\  
\hline
\\
\end{tabular}
\label{table:MF25_improvements_clusters}
\end{table*}

\begin{figure*}[t!]
\centering
\includegraphics[height=7.1cm]{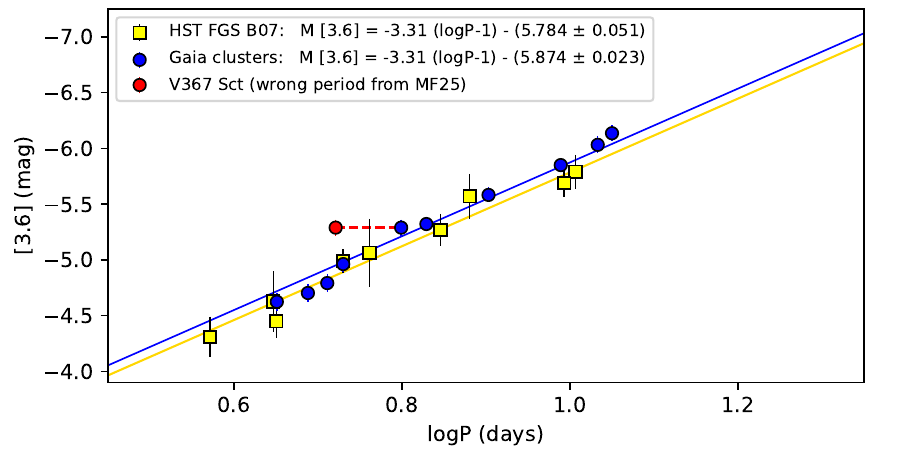} 
\caption{P--L relation for Milky Way cluster Cepheids (blue) as selected by \citetalias{MF2025}, after applying the corrections listed in Table~\ref{table:MF25_improvements_clusters}, and for the HST FGS sample from \citetalias{Benedict2007} (yellow). The red point shows V367 Sct as plotted using \citetalias{MF2025}'s incorrect period. The blue version uses the correct $\log P = 0.799$ value. \\  }
\label{fig:PL_outlier}
\end{figure*}

\subsection{Reproducing the initial measurement by \citetalias{MF2025}}
\label{sec:remake_MF25}

In this section, we adopt the same data sets and assumptions as in \citetalias{MF2025} to attempt a reproduction of the reported results. That study identified 13 cluster Cepheids in the Milky Way based on common proper motions of member stars. Although a much larger sample of $>30$ cluster Cepheids is available in the literature \citep[e.g.][]{Anderson2013, Breuval2020, Riess2022b, CruzReyes2023, Wang2024}, the analysis in \citetalias{MF2025} is carried out in the \textit{Spitzer} [3.6 $\mu$m] filter in order to minimize the effects of interstellar absorption and width of the instability strip. In that filter, only 37 Milky Way Cepheids have available photometry from \cite{Monson2012}, among which only 30\% are in clusters, thereby significantly limiting the usable sample size. Among the 13 cluster Cepheids retained in \citetalias{MF2025}, EV Sct and CS Vel were explicitly excluded and do not appear in their Figure 3, therefore only 11 cluster Cepheids are used. In \citetalias{MF2025}'s Table 1, the Cepheid WZ Sgr is associated with the open cluster ``Turner 2''. \citetalias{MF2025} note in their Appendix A.9. that ``{\it the reality of Turner 2 as a bona fide cluster is questionable}". This cluster appears in \cite{Hunt2023} in the list of clusters reported in the literature that were not recovered using {\it Gaia} EDR3. Similarly, \citet{CruzReyes2023} performed a search for clusters in the vicinity of Cepheids and did not recover a cluster associated with WZ Sgr based on {\it Gaia} EDR3 astrometry. Turner 2 is listed in the \cite{Kharchenko2013} catalog but with only one star used to measure the radial velocity, and only 24 stars within a $r_1$ radius. It does not exist in other modern catalogs based on {\it Gaia} \citep{CantatGaudin2020, CastroGinard2022}, is mentioned in \cite{Turner1993} under a different name (``C1814-191a"), and is described as a ``sparse cluster". Since we cannot reproduce the mean parallax of this cluster with \textit{Gaia} EDR3 parallaxes, we exclude it from our reanalysis. 
Following \citetalias{MF2025}, we fix the P--L slope to the LMC value \citep[-3.31,][for $6 < P < 60$ days]{Scowcroft2012} and we fit the P--L relation for the  cluster Cepheid sample (N=10) from \citetalias{MF2025}, where the absolute magnitudes $M_{[3.6]}$ are:
\begin{equation}
M_{[3.6]} = m_{[3.6]} -0.203 \, E(B-V) - (m-M)_{0, \, \rm corr}
\end{equation}
Apparent magnitudes $m_{[3.6]}$ are from Table 4 in \cite{Monson2012}, the distance moduli $(m-M)_{0, \, \rm corr}$ for each Cepheid are from Table 1 in \citetalias{MF2025} and already include their 0.26 mag offset, and $E(B-V)$ are from \citetalias{MF2025} as well. A systematic error of 0.016 mag is included for the photometry error in $m_{[3.6]}$, and 0.06 mag errors are added in quadrature to absolute magnitude errors for the intrinsic width of the instability strip. We thus obtain:
\begin{equation}   
M_{[3.6]} = -3.31 \, (\log P -1) - (5.823 \pm 0.022) 
\label{eq_B}
\end{equation}
with $\sigma = 0.071 \, \rm mag$. This corresponds to the initial cluster Cepheid P--L relation as obtained in \citetalias{MF2025}, with the same pivot period ($\log P_0  = 1$). We can compare it with that in the LMC to derive a first estimate of the metallicity dependence $\gamma$. In the LMC, using data from \cite{Scowcroft2012} with $E(B-V) = 0.07 \pm 0.01 \, \rm mag$, $P<60 \, \rm days$, a fixed slope of $-3.31$ and a geometric distance to the LMC of $d = 49.59 \pm 0.09 \pm 0.054 \, \rm kpc$  \citep{Pietrzynski2019}, we obtain:
\begin{equation}
M_{[3.6]} = -3.31 \, (\log P-1) - (5.787 \pm 0.010) 
\label{eq_LMC_M}
\end{equation}
The difference in P--L intercept ($\Delta \beta$) between the Milky Way clusters (Eq.~\ref{eq_B}) and the LMC sample (Eq.~\ref{eq_LMC_M}) is $0.036 \pm 0.024 \, \rm mag$. For the Milky Way and LMC Cepheid metallicities, we adopt [Fe/H] = $+0.146 \pm 0.075$ and $-0.407 \pm 0.02 \, \rm dex$ respectively \citep{Breuval2022, Romaniello2021}. The metallicity effect $\gamma$ is therefore defined as:
\begin{equation}
\gamma = \frac{\Delta \beta}{\Delta \rm [Fe/H]} 
\label{eq_gamma}
\end{equation}
and yields $\gamma = -0.06 \pm 0.05 \, \rm mag/dex$. As \citetalias{MF2025} conclude, this method suggests that the metallicity dependence of the P--L intercept is small and not significantly negative, contrary to what most empirical studies find. Instead of adding a 0.26 mag offset to match an expected dispersion, if \citetalias{MF2025} had applied a slightly larger magnitude offset to exactly match the HST FGS parallax P--L, a metallicity dependence $\gamma \sim 0$ would have been obtained. In the next section, we re-evaluate the Milky Way Cepheid P--L relation and the metallicity dependence using better (see Table~\ref{table:MF25_improvements_clusters}) and more modern data.   \\

\subsection{Updating the \citetalias{MF2025} measurement}

Here, we adopt the same Cepheid sample as in \citetalias{MF2025} with their \textit{Gaia} EDR3 parallaxes taken from \cite{Riess2022b} \citep[or see][]{CruzReyes2023, Wang2024} instead of the distance moduli with ad hoc calibration from \citetalias{MF2025}. These parallaxes include the \citet{Lindegren2021_plx_bias} correction with no additional terms. \citetalias{MF2025} claim that clusters ``{\it improve statistical precision by a factor of 16}'' compared to individual Cepheids. However, it should be noted that clusters only improve parallax precision by a factor of $\sim$3, after accounting for an additional $\sim7~ \mu$as term in the parallax error due to angular covariance \citep{MaizApellaniz2021, Vasiliev2021}. These are included in the mean cluster parallax uncertainties. After inspecting the P--L diagram, we noticed that the cluster Cepheid V367 Sct (shown in red in Fig.~\ref{fig:PL_outlier}) appears as an outlier compared to the other well aligned data points, and that it was excluded from the P--L figures in \citetalias{MF2025} without justification. For this Cepheid, \citetalias{MF2025} adopt $\log P$=0.721 (see their Table 1) as in \citet{Madore1975}. However, \cite{Madore1978} identify V367 Sct as a double-mode Cepheid and provide fundamental and first overtone periods of 6.29307 days ($\log P$ = 0.799) and 4.38466 days ($\log P$ = 0.642), respectively. All recent studies use $\log P$=0.799 as well \citep{Monson2012, CruzReyes2023, Breuval2020, Riess2022b, ZhouChen2021}, which is the correct fundamental-mode pulsation period. In the following, we update the pulsation period of this Cepheid with $\log P$=0.799.

\begin{figure*}[t!]
\centering
\includegraphics[height=12.4cm]{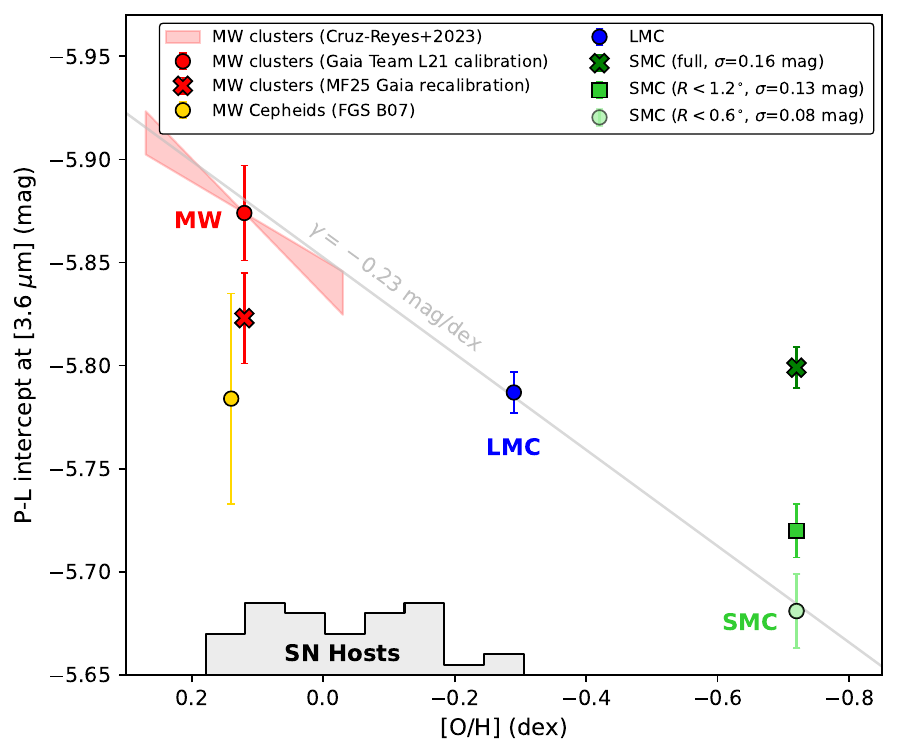} 
\caption{P--L intercept $\beta$, where $M = \alpha \, (\log P-1) + \beta$, in the [3.6 $\mu$m] filter for Milky Way cluster Cepheids (red), the sample of 10 bright Cepheids with HST FGS parallaxes from \citetalias{Benedict2007} (yellow), LMC Cepheids (blue) and Cepheids in different regions of the SMC (green). The red "X" shows the P--L intercept obtained after the \citetalias{MF2025} recalibration of {\it Gaia} EDR3 parallaxes. The metallicity on the horizontal axis is expressed in [O/H] for a direct comparison with abundances in SNIa host galaxies (grey histogram). The slope of the grey line represents a $\gamma \sim -0.23$ mag/dex metallicity dependence obtained by fitting a straight line through the MW, LMC and SMC. Conversely, the $\gamma \sim 0$ result by \citetalias{MF2025} is obtained from a comparison between the MW red ``X'' marker, the LMC, and the SMC dark green ``X'' marker, which align with a horizontal line.     \\  }
\label{fig:intercept_OH}
\end{figure*}

The successive updates presented here are shown in Fig.~\ref{fig:intercept_OH} and are listed in Table~\ref{table:MF25_improvements_clusters}, where the first row is the original P--L relation obtained in the same conditions as \citetalias{MF2025}. Our best re-evaluation of the Milky Way cluster Cepheid P--L relation is:
\begin{equation}
M_{[3.6]} = -3.31 \, (\log P -1) - (5.874 \pm 0.023)
\label{eq_R}
\end{equation}
with $\sigma = 0.080 \, \rm mag$. It is represented in Fig.~\ref{fig:intercept_OH} with a red circle. Assuming the same metallicities as in Sect.~\ref{sec:remake_MF25} and the same LMC P--L intercept of $-5.787 \pm 0.010$ mag, we now obtain:
\begin{equation}
\gamma_{[3.6]} = -0.16 \pm 0.05 \, \rm mag/dex,
\end{equation}
in good agreement with most estimates from the literature. The key difference here is a net, mean 0.05 mag difference between the \citetalias{MF2025} and \citet{Lindegren2021_plx_bias} calibration of {\it Gaia} parallaxes (see $\beta$ values in Table~\ref{table:MF25_improvements_clusters}) which represents about half the size of the metallicity effect between the Milky Way (MW) and the LMC (or the full size if the HST FGS parallaxes are assumed to define the reference).

We further note that \citetalias{MF2025} claim that the results from \cite{Breuval2021, Breuval2022} must be ``{\it extrapolated into the metallicity range covered by most of the more distant Type Ia supernova that calibrate host galaxies}''. However, the analysis presented in \cite{Breuval2021, Breuval2022} covers the Milky Way, the LMC and the SMC. As shown in Fig.~\ref{fig:intercept_OH}, the metallicity of Milky Way Cepheids is similar to (if not higher than) that of SNIa hosts, so there is no extrapolation needed. \\

\subsection{A differential calibration of the metallicity effect between the LMC and the SMC} 

An independent approach for measuring the metallicity dependence $\gamma$ comes from the comparison of Cepheids in the LMC and Small Magellanic Cloud (SMC) versus their geometric distance difference determined from DEBs. Section 4.2 of \citetalias{MF2025} revisits this purely differential calibration of $\gamma$ based on the metallicity difference between the LMC and SMC, and the difference in P--L intercept between the two galaxies in the [3.6 $\mu$m] filter. The analysis follows the approach of \cite{Wielgorski2017}, 
which yielded $\gamma \sim 0$ mag/dex in the $V$, $I$, $J$, $H$, and $K$ filters. However, the calibration of Cepheids with the geometric DEB distance is complicated by the well-known and substantial depth of the SMC \citep{Ripepi2017, Scowcroft2016, Subramanian2015, Jacyszyn2016}. \citet{Breuval2022, Breuval2024} analyzed the SMC P--L intercept and scatter as a function of the separation from the SMC core, found strong evolution of both, and concluded that these effects, attributed to the SMC's elongated shape, strongly affected the \cite{Wielgorski2017} result: applying a geometry correction (e.g. derived by \cite{Graczyk2020} from the planar distribution of the eclisping binaries) to each Cepheid depending on their position, as well as using only Cepheids in the SMC 0.6 degree core-region yields $\gamma \sim -0.15$ mag/dex, in good agreement with other empirical studies.   


Here, we follow the same approach with the LMC and SMC Cepheid samples and the [$3.6 \, \mu$m] photometry from \citet{Scowcroft2011, Scowcroft2016} respectively. We assume $E(B-V) = 0.07 \pm 0.01$ mag in the LMC and $E(B-V) = 0.03 \pm 0.01$ in the SMC \citep{Skowron2021} and we select Cepheids with $P < 60$ days. In the SMC we use [Fe/H] = $-0.785 \pm 0.085 \, \rm dex$ from Romaniello, M. et al., (in prep.). With a slope fixed to the LMC value of $\alpha = -3.31$, the P--L intercepts in the LMC and SMC are $\beta = -5.787 \pm 0.010$ and $\beta = -5.790 \pm 0.010$ mag respectively. From Eq.~\ref{eq_gamma}, we obtain an initial estimate of $\gamma = +0.008 \pm 0.037 \, \rm mag/dex$, in agreement with \citetalias{MF2025}'s claim. To account for the depth effects in the SMC, we now apply the geometry correction from \cite{Breuval2024} and we limit the sample to the SMC core region. The P--L intercept in the SMC and the resulting metallicity dependence $\gamma$ are listed in Table~\ref{table:LMC_vs_SMC} for the full sample, for $R<1.2^{\circ}$ and $R<0.6^{\circ}$, with and without the geometry corrections. The P--L intercepts obtained with geometry corrections are shown in Fig.~\ref{fig:intercept_OH} in green. We see that the use of the SMC geometry and core region, {\it independently corroborated} by the tighter P--L dispersion, produce a value of $\gamma \sim -0.3$ mag/dex, in better agreement with the canonical $\gamma \sim -0.2 \, \rm mag/dex$ from the literature. Finally, we note that \cite{Wielgorski2017} was based on the old SMC geometric distance by \cite{Graczyk2014}, and that this measurement has been updated in \cite{Graczyk2020} with a larger sample of 15 eclipsing binaries. However, none of the recent studies by \cite{MF2024a} and \cite{Madore2025_Iband} use the modern value from \cite{Graczyk2020}. This small difference of 0.027 mag in SMC distance modulus between \cite{Graczyk2014} and \cite{Graczyk2020} might affect $\gamma$ at the $\sim 0.07 \, \rm mag/dex$ level (assuming a 0.38 dex difference between the LMC and SMC). 

To summarize, the case by \citetalias{MF2025} for $\gamma \sim 0$ between the LMC and the SMC requires 1) the use of an earlier DEB result from \cite{Graczyk2014} rather than the more recent \cite{Graczyk2020}, 2) neglecting the SMC geometry described in \cite{Graczyk2020} and \cite{Breuval2024}, and 3) neglecting strong and independent evidence that the SMC has significant depth effects which are largely if not fully resolved by limiting to the SMC core and application of the geometric model. \\

\begin{table*}[t!]
\centering
\caption{P--L intercept $\beta$ in the [3.6 $\mu$m] filter for the SMC sample from \cite{Scowcroft2016} and resulting metallicity dependence $\gamma$ from the differential calibration with respect to the LMC. We present the results for different regions around the SMC center, and with/without including geometry corrections. The P--L slope is fixed to $-3.31$ mag/dex. The initial calibration (full sample, no correction) and final result ($R<0.6^{\circ}$, with corrections) are both highlighted in bold.  \\}
\begin{tabular}{c c c c c l}
\hline
\hline
Geometry  & Param. & $R<0.6^{\circ}$ & $R<1.2^{\circ}$ & Full  \\
correction           &        & (N = 21)        &  (N = 41)       & (N = 85) \\
\hline
    & $\beta_{\rm SMC}$ & $-5.665 \pm 0.016$ & $-5.712 \pm 0.011$ & $-5.790 \pm 0.010$       & (mag)   \\  
No  & $\gamma$          & $-0.323 \pm 0.090$ & $-0.198 \pm 0.063$ & {\bf +0.008 $\pm$ 0.037} & (mag/dex)  \\ 
    & $\sigma_{\rm PL}$ & $0.089$            & $0.131$            & $0.159$                  & (mag) \\
\hline
    & $\beta_{\rm SMC}$ & $-5.673 \pm 0.016$         & $-5.693 \pm 0.011$ & $-5.674 \pm 0.010$ & (mag)    \\  
Yes & $\gamma$          & {\bf $-$0.302 $\pm$ 0.086} & $-0.249 \pm 0.072$ & $-0.299 \pm 0.079$ & (mag/dex)    \\  
    & $\sigma_{\rm PL}$ & $0.078$                    & $0.127$            & $0.159$            & (mag) \\
\hline
 \\
\end{tabular}
\label{table:LMC_vs_SMC}
\end{table*}

\subsection{TRGB versus Cepheid Distance Moduli} 

Sect.~6 of \citetalias{MF2025} presents a comparison between Cepheid and Tip of the Red Giant Branch (TRGB) distances across a large sample of galaxies. Based on the absence of a significant trend with host metallicity, the study concludes that $\gamma =0$. As previously discussed \citep[see][Sect. 4.3]{Breuval2024}, this approach presents challenges, as TRGB distances are not necessarily free from metallicity dependence \citep{Rizzi2007, Koblischke2024}. The TRGB data used in \citetalias{MF2025} are drawn from disparate sources in the literature, despite the availability of recent and more uniformly measured samples \citep[e.g., the Extragalactic Distance Database;][]{Tully2009, Anand2021}, which could potentially affect the outcome. Moreover, much of the leverage in the TRGB–Cepheid comparison comes from a small number of extremely metal-poor galaxies with sparse Cepheid populations. These environments are rare in terms of spatial density, and the assumption that the few Cepheids present share the low average metallicity of their hosts may not be robust. It is more likely that these are high metallicity spots with some recent star formation, making it inappropriate to assume uniform low metallicity. In fact, the metallicities inferred for these Cepheids, around [O/H]$\sim -1.2$ dex, are lower than those found from direct measurements. For example, for the metal-poor galaxy Sextans A, \citetalias{MF2025} assume 12 + $\log$(O/H) = 7.49 dex from \cite{Sakai2004}, equivalent to [O/H] = $-1.2$ dex. On the other hand, \cite{Kaufer2004} find [Fe/H] = $-0.99$ dex from high resolution spectroscopy of supergiants, which is equivalent to [O/H] $\sim -0.93$ dex (the relationship [O/H] $\sim$ [Fe/H] + 0.06 dex is adopted from \citealt{Riess2022a}), significantly less metal-poor. In contrast, direct geometric and spectroscopic measures of the most metal-poor Cepheids in the MW from the C-MetaLL project \citep{Trentin2024} result in negative values of $\gamma$. Furthermore, \citet{Bhardwaj2024} used new homogeneous photometry and high-resolution spectroscopic metallicities for 61 Milky Way Cepheids and found that $\gamma$ is more negative for metal-poor Cepheids ($-1.1 <$ [Fe/H] $< -0.3$), which are more distant and have larger parallax uncertainties. The quality of the Cepheid data and the number of Cepheids in the metal poor hosts used in \citetalias{MF2025} is well-below the measures in the better-studied LMC, SMC, and Milky Way. It is beyond the scope of this paper to investigate whether the abundances used in \citetalias{MF2025} are representative of the Cepheids in the low-metallicity galaxies. However, it is important to note that mean abundances that apply to TRGB measurements may not be adequate for Cepheids, which are generally much younger. A further complication that can arise when going to very low-metallicity regime is that the [$\alpha$/Fe] ratios are no longer consistent between the samples, with $\alpha$-enhancement simulating a higher metallicity \citep{Salaris1993}. This could have a significant impact on opacities, and hence, on a metallicity effect specified only relative to iron.


A specific example of this approach can be found in Section 4.2.2 of \citetalias{MF2025}, where a comparison of Cepheid \citep[$\mu = 24.29 \pm 0.03$ mag,][]{Freedman2009}, TRGB \citep[$\mu = 24.30 \pm 0.03$ mag,][]{Hatt2017}, and RR Lyrae \citep[$\mu = 24.28 \pm 0.04$ mag,][]{Hatt2017} distance moduli to the metal-poor galaxy IC1613 is presented. From the close agreement between the three values, they conclude that the metallicity dependence is consistent with zero. However, the \citetalias{MF2025} sample, described as ``{\it well populated Cepheid P--L relations}", is actually based on only 5 Cepheids and each with single, random epoch observations in the {\it Spitzer} filters from \cite{Freedman2009}. Given the intrinsic dispersion of $\sim0.08$ mag in the {\it Spitzer} P--L relation, the quoted uncertainty of $\pm0.03$ mag in the Cepheid distance modulus may be underestimated. Moreover, a broader set of distance measurements in the literature, such as $\mu = 24.20 \pm 0.07$ mag \citep{Udalski2001}, $\mu = 24.29 \pm 0.03$ mag \citep{Pietrzynski2006}, $\mu = 24.38 \pm 0.05$ mag \citep[][TRGB]{Jacobs2009}, and $\mu = 24.31 \pm 0.06$ mag \citep[][RR Lyrae and Cepheids]{Dolphin2001}, suggests a wider spread among different estimates of the distance to IC1613 than that considered in \citetalias{MF2025}. While these differences are mostly within $1–2\sigma$, they highlight the importance of accounting for measurement limitations and sample size when drawing conclusions about metallicity effects.

We further note that Fig.~20 in \citetalias{MF2025} highlights the \cite{Breuval2022} results in red and yellow, and the arrow near the Milky Way represents the \citetalias{MF2025} parallax correction (in magnitude space). In this figure taken from \cite{MF2024a} but later updated in \citet{MF2024b}, \citetalias{MF2025} combine observations made in multiple wavelengths (the filter used to produce this plot is not specified), which invalidates the method due to inconsistent wavelength use. While the \citetalias{MF2025} sample is described overall as ``{\it well observed galaxies}'', P--L relations often have $<30$ Cepheids and many of the metallicity measures are imprecise compared to the well-established values in the Milky Way, LMC and SMC. Cepheid data are from 21 different sources, including before HST spherical aberration was fixed. The same trend can be found throughout the rest of the \citetalias{MF2025} paper (see Table~\ref{table:choice_data} in the Appendix). For example, in Figure 21 in \citetalias{MF2025}, Cepheids in M33 come from a small sample of ground-based data from \cite{Freedman1991} rather than 154 Cepheids with HST observations in \cite{Breuval2023}. In Figures 22 and 23 in \citetalias{MF2025}, residuals from extinction curve fits in SH0ES galaxies are taken from \cite{Riess2016} instead of the latter data release from \cite{Riess2022a}. For the DEBs geometric distance to the SMC, \citetalias{MF2025} cite \cite{Graczyk2014} and ignore the update by \cite{Graczyk2020}. In contrast, the MW, LMC, and SMC Cepheids now have direct, spectroscopic metallicity measures, direct geometric distance measures, and uniform, high quality photometry from HST WFC3. It is therefore not surprising that inhomogeneous data wash out the signal that is seen with modern measurements. It is not clear why we should be limited to extracting results from such past data when there are more modern, better data sets available. \\

\subsection{Residuals from the Milky Way Cepheid Sample} 

In Section 5, \citetalias{MF2025} report that no correlation between P--L residuals and metallicity is found, and characterize this conclusion as ``{\it not new}", citing \cite{Narloch2023} in support. However, this interpretation does not align with the conclusions explicitly stated in \cite{Narloch2023}. While that study does note the absence of a clear trend between residuals and metallicity in any band, including reddening-free Wesenheit analogs, it also emphasizes that the underlying metallicity range is narrow, limiting the ability to detect any potential trend. \cite{Narloch2023} state:
``{\it The comparison of the P--L/P--W relations residuals with metallicity does not show any clear trends in any band, but it should be borne in mind that the range of metallicity used is rather narrow, and trends could possibly be revealed for a wider range of metallicities. Because of the small range of the metallicities of our Cepheids, we also decided not to investigate a period–luminosity–metallicity (PLZ) relation, however, such an analysis will be the subject of future projects.}" Additionally, the scatter values quoted in Table 3 of \citetalias{MF2025}, attributed to \cite{Breuval2021}, do not match the values published in that work or in any of the related Breuval et al. papers, and also conflict the values from \cite{Madore2012} (see Table \ref{table:PL_scatter_B21}). Therefore, the validity of these claims is questionnable.  \\

\begin{table}[t!]
\centering
\caption{P--L relation scatter (in mag) quoted in \citetalias{MF2025}, compared to the actual published values in \cite{Breuval2021}, \cite{Breuval2022} and \cite{Madore2012}. Values quoted by \citetalias{MF2025} do not match the published ones and are systematically larger, making them look worse than they are. \\  }
\begin{tabular}{c c c c c}
\hline
\hline
Band     & MF25 & B21 & B22 & MF12   \\
 & (Table 3) & (Table 1) & (Table 4) & (Table 1)  \\
\hline
$V$ & 0.27 & 0.25 & 0.22 & 0.27 \\
$I$ & 0.26 & 0.23 & 0.19 & 0.18 \\
$J$ & 0.23 & 0.18 & 0.19 & 0.14 \\
$H$ & 0.22 & 0.17 & 0.18 & 0.12 \\
$K$ & 0.20 & 0.17 & 0.17 & 0.11 \\
\hline
\end{tabular}
\label{table:PL_scatter_B21}
\end{table}

\begin{table*}[t!]
\centering
\caption{Summary: Consequences of the different treatments of {\it Gaia} EDR3 parallaxes. \\   }
\begin{tabular}{l c c}
\hline
\hline
& {\it Gaia} Collaboration \citepalias{Lindegren2021_plx_bias} & \citet{MF2025} \\
\hline
Distance to the Pleiades Cluster & 135 pc & 120 pc \\
\hline
Parallax of Distant Quasars & $\sim 0 \, \mu$as & $\sim -25 \, \mu$as \\
\hline
Inferred value of $H_0$ & 73 \hunit & 74.3 \hunit \\
\hline
~ \\
\end{tabular}
\label{table:consequences}
\end{table*}

\subsection{The Metallicity Dependence from Baade–Wesselink P--L Relations}

In their section 4.1, \citetalias{MF2025} discuss the results by \citet{Fouque2007} and \citet{Storm2011b} based on Baade-Wesselink (BW) distances of Cepheids in the MW, LMC and SMC. The latter work resulted in a small negative metallicity effect, which within errorbars can be interpreted as consistent with zero. However, a further update of these two works is presented in \citet{Gieren2018}, based on an extended sample of Cepheids in the SMC (from N=5 to N=31 Cepheids), and was not mentioned in \citetalias{MF2025}. \citet{Gieren2018} used the exact same method as in \citet{Storm2011b} and obtained a metallicity dependence between $-0.22$ and $-0.33$ mag/dex for optical and near-infrared bands as well as $W_{VI}$ and $W_{JK}$ Wesenheit indices, with errors of $\sim0.15$ mag/dex in optical passbands and $<0.1$ mag/dex for infrared bands and Wesenheit indices, which is fully consistent with the canonical $-0.2$ mag/dex value. 

In the BW method \citep{Baade1948, Wesselink1946}, the changes in the apparent angular diameter of the star, inferred from photometry and surface brightness-color relations, are compared to the physical radius displacement obtained from the integration of the velocity of the stellar atmosphere. The main source of uncertainty for BW distances is the value of the projection factor ($p$-factor) used to translate radial velocities measured from spectra into the true velocity of the stellar atmosphere. The precise value of this parameter, as well as its possible dependence with any physical parameter of the star, are still open questions \citep[see][and references therein]{Trahin2021}.

The large P--L scatter obtained from BW distances in \citet{Storm2011b} and \citet{Gieren2018} is indeed high, as noted by \citetalias{MF2025}, and does not decrease with wavelength as expected. However, it does not rule out the BW technique as unreliable, but rather indicates that the spread of the P--L relations is dominated by statistical errors on individual distance measurements, related not only to the poor knowledge of the $p$-factor, but also to the quality of photometry and radial velocities used in the analysis. In addition, as BW distances linearly depend on the assumed $p$-factor, any systematic shift of this parameter will not influence the measured relative distance modulus of the MW, LMC and SMC Cepheids, which is the basis for the \citet{Gieren2018} calibration of the metallicity effect. Further, \citet{Nardetto2011} showed that the $p$-factor is largely independent of metallicity and \citet{Thompson2001} showed that the eﬀect on surface brightness relations is also very small. Even though the BW method is not competitive with {\it Gaia} EDR3 parallaxes, the LMC distance modulus obtained by \citet{Storm2011b} and \citet{Gieren2018} is in a excellent agreement with the geometric measurement from eclipsing binaries \citep{Pietrzynski2019}. Summing this up, regardless of the exact value (or period-dependence) of the $p$-factor, the results obtained from the BW method contradict the conclusions of \citetalias{MF2025} and support a significant negative effect in the optical and near-infrared. \\

\section{Consequences of the \citetalias{MF2025} Re-calibration of {\it Gaia} Parallaxes}
\label{sec:consequences}

The {\it Gaia} team, following \citetalias{Lindegren2021_plx_bias}, recommends a three-step calibration procedure: use {\it Gaia} EDR3 parallaxes, apply the \citetalias{Lindegren2021_plx_bias} correction as a function of magnitude, color, and sky position, and, if applicable, solve for an additional additive offset tailored to specific source populations like Cepheids. In contrast, \citetalias{MF2025} bypasses the \citetalias{Lindegren2021_plx_bias} correction and introduces a fundamentally different, multiplicative correction that lacks justification within {\it Gaia}’s parallax measurement framework. While this approach may yield similar results for cluster Cepheids, it will diverge significantly at other distances. Although \citetalias{MF2025} do not explicitly state that their recalibration of {\it Gaia} parallaxes should be applied outside of their Cepheid sample, we demonstrate that applying the \citetalias{MF2025} correction to other sources, such as the Pleiades or quasars, leads to inaccurate results, underscoring its lack of general validity. In particular, the Pleiades are cluster stars, and therefore the \citetalias{MF2025} correction, derived for cluster Cepheids, would be expected to apply there. The consequences of the different treatments of {\it Gaia} parallaxes discussed in this paper are summarized in Table~\ref{table:consequences}. \\

\subsection{Distance to the Pleiades}

\begin{figure}[t!]
\centering
\includegraphics[height=6.1cm]{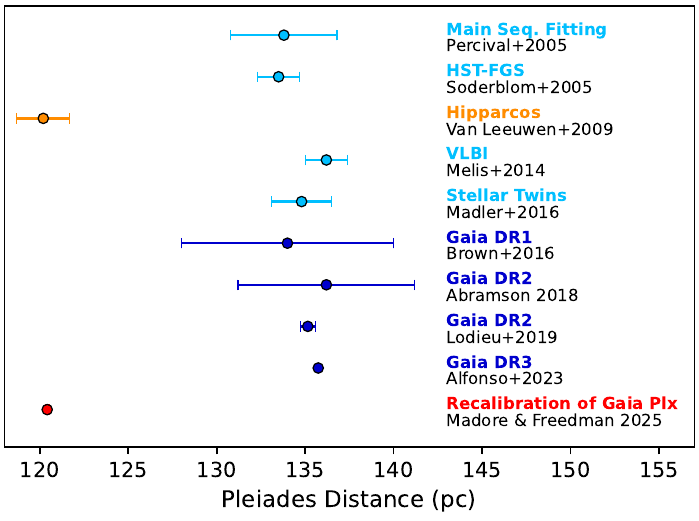} 
\caption{Distance estimates to the Pleiades open cluster. Light blue values are pre-{\it Gaia} measurements, orange is Hipparcos, dark blue are {\it Gaia}-based distances and the red point corresponds to {\it Gaia} EDR3 after applying the recalibration suggested by \citetalias{MF2025}.   }
\label{fig:pleiades}
\end{figure}

The Pleiades is a very nearby open cluster, extensively studied in the Milky Way and much like the open clusters which host Cepheids farther away, so presumably the \citet{MF2025} approach would apply to all open clusters. Here, we evaluate the impact of the \citetalias{MF2025} recalibration of {\it Gaia} cluster parallaxes for the Pleiades cluster distance. We recall that the \citetalias{MF2025} recalibration relies on an additive 0.26 term in distance modulus, equivalent to a multiplicative correction in parallax space (Eq.~\ref{eq_MF_dm} and \ref{eq_MF_plx}). In the early 2000s, the Hipparcos astrometric mission measured a distance of 120 pc to the Pleiades \citep{VanLeeuwen2009}, which was proven to be too short by about 15 pc thanks to more precise measurements, including the recent {\it Gaia} parallaxes \citep{Brown2016, Abramson2018, Lodieu2019, Alfonso2023}. In addition, other methods independent from {\it Gaia} have confirmed the higher distance of about 135 pc using main sequence fitting \citep{Percival2005}, HST-FGS parallaxes \citep{Soderblom2005}, very long baseline radio interferometry \citep{Melis2014} or stellar twins \citep{Madler2016}. Applying the \citetalias{MF2025} recalibration to {\it Gaia} EDR3 parallaxes for the Pleaides lowers their distance to $\sim$ 120 pc (red point in Fig.~\ref{fig:pleiades}), similar to the Hipparcos measurement, and 15 pc shorter than the canonical value.   It is interesting to note that the \citetalias{MF2025} reliance on the HST FGS parallaxes to recalibrate {\it Gaia} EDR3 would produce conflicting results with the HST FGS measured parallax of the Pleiades, indicating that a multiplicative recalibration of {\it Gaia} parallaxes is inconsistent with a wider sample of HST FGS parallax measures. In fact, the \citetalias{Lindegren2021_plx_bias} additive calibration would produce far better agreement to both sets of HST FGS parallaxes. \\

\begin{figure}[t!]
\centering
\includegraphics[height=5.7cm]{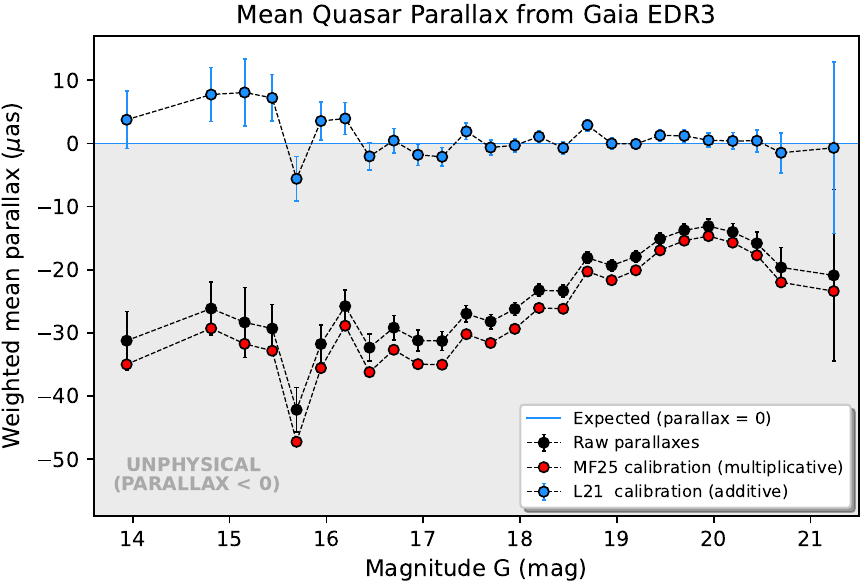} 
\caption{Mean parallaxes of distant quasars as a function of {\it Gaia} $G$ magnitude from the {\it Gaia} EDR3 catalog without any correction (black), with the multiplicative correction proposed by \citetalias{MF2025} (red) and with the additive correction recommended by the {\it Gaia} team \citepalias[][in blue]{Lindegren2021_plx_bias}. Quasar parallaxes are expected to be equal to zero. Figure adapted from \citetalias{Lindegren2021_plx_bias}.   }
\label{fig:quasars}
\end{figure}

\subsection{Non-zero Parallaxes of Quasars}

The original {\it Gaia} EDR3 parallaxes are affected by a small systematic bias \citepalias{Lindegren2021_plx_bias} that varies across the sky and depends on magnitude, color and position. On average, this bias is of the order of $\sim 20 \, \rm \mu as$ for quasars. This can be seen, for example, in the negative (hence unphysical\footnote{Negative {\it Gaia} parallaxes can be caused by errors in the observations. Even if a negative distance has no physical meaning, a certain number of stars are expected to have negative parallaxes from an error propagation perspective. The negative parallax tail is a useful diagnostic for the quality of the astrometric solution \citep{Arenou2018, Lindegren2021_astrometric_solution}.}) parallaxes measured for distant quasars, whose true parallaxes must be zero (see black points in Fig.~\ref{fig:quasars}). Applying the \citetalias{Lindegren2021_plx_bias} correction as an additive term to the parallaxes brings them back to zero within $\sim1 \, \mu$as (blue points in Fig.~\ref{fig:quasars}). In contrast, the recalibration proposed by \citetalias{MF2025} multiplies {\it Gaia} EDR3 parallaxes by $\sim1.12$, which makes them even more negative -- amplifying but not correcting the small, negative quasar parallaxes. The \citetalias{MF2025} recalibration results in a systematic bias that cannot be reconciled with the known zero-parallax of quasars.  \\

\subsection{The Hubble Constant}

The main result of the Key Project was a value of 72 \hunit for the Hubble constant, based on the LMC as only anchor and assuming $\gamma = -0.2$ mag/dex for the Cepheid metallicity dependence \citep{Freedman2001}. In that case, because the LMC is more metal-poor than SNIa host galaxies, the sensitivity of $H_0$ to the metallicity correction is important. \cite{Freedman2001} write: ``{\it The effect is systematic [..] if no correction for metallicity is applied, the value of $H_0$ is increased by $\sim$4\% ($\sim$3 \hunit)}". However, as stated in the introduction, the impact of the Cepheid metallicity dependence $\gamma$ for $H_0$ is now limited, thanks to the similarity in metallicity between Cepheids in anchor galaxies and in SNIa hosts. Although the LMC and SMC are particularly metal-poor, the use of the metal-rich Milky Way and NGC 4258 balances this difference \citep[see Fig. 21 in][]{Riess2022a}.

The \citetalias{MF2025} recalibration of {\it Gaia} parallaxes applied to Cepheids in the Milky Way brings their P--L relation closer to that of the HST FGS \citetalias{Benedict2007} sample by 0.04 mag. 
This change results in a less negative (i.e. larger, fainter) P--L intercept for {\it Gaia} Milky Way Cepheids, implying shorter distances to SNIa host galaxies, and thus a higher value for the Hubble constant. The \citetalias{MF2025} correction would raise $H_0$ from the \cite{Riess2022b} {\it Gaia} calibration to 74.4 km/s/Mpc. 
Similarly, \cite{Riess2016} used the HST FGS parallaxes from \citetalias{Benedict2007} to calibrate $H_0$, resulting in 76 km/s/Mpc, before the replacement of these with {\it Gaia} parallaxes \citep[see Appendix B2 in][]{Riess2022a}. On the other hand, more recent distance ladders that include more metal-rich anchors like Gaia clusters \citep{Riess2022b} change the anchor-to-calibrator weighting to metal-rich so that $\gamma$=0 can slightly lower H$_0$ by 0.4 km/s/Mpc \citep{H0DN2025}.  \\


\section{Discussion}
\label{sec:discussion}

In this paper, we revisited the recalibration of {\it Gaia} EDR3 parallaxes proposed in \citetalias{MF2025}. We first reproduced the cluster Cepheid P--L relation as presented in that work, and then replaced the 0.26 mag magnitude offset proposed by \citetalias{MF2025} (equivalent to a multiplicative parallax correction) with the additive parallax offset $\varpi_{\rm L21}$ recommended by the {\it Gaia} team \citep{Lindegren2021_plx_bias}. This substitution changes the Cepheid metallicity dependence $\gamma$ from $\sim 0$ to $-0.16$ mag/dex, bringing it into close agreement with both widely adopted empirical values and predictions from stellar models. We also examined additional methods employed in \citetalias{MF2025}, including the comparison of P--L relations in the LMC and SMC. Using an updated treatment of the SMC geometry and limiting the Cepheid sample to the SMC core region, we find that this approach likewise supports a negative metallicity dependence. We further note limitations in the use of Cepheid–TRGB distance comparisons as a test of $\gamma$, which we show does not provide a reliable test of the metallicity dependence at the fidelity of what can be directly measured from the best LMC, SMC and MW data. Finally, we assess the broader implications of the \citetalias{MF2025} {\it Gaia} parallax recalibration and find that it leads to several inconsistencies: notably, a distance many $\sigma$ too short for the Pleiades cluster, and negative and unphysical parallaxes for extragalactic quasars. 

It is likely that the upcoming {\it Gaia} DR4, based on several more years of data and with improvements in the focal plane calibration, will reduce or remove the dependence of the {\it Gaia} parallax offset on stellar properties and location, thereby obviating the need for an additional parallax correction for bright Cepheids. This would result in a significant improvement in the calibration of the Period-Luminosity relation for Milky Way Cepheids, and a reduction of any residual systematic uncertainties.

The metallicity dependence is a small effect that has a mild impact on the value of the Hubble constant. As it remains difficult to measure its value, a precise determination therefore requires using the highest-quality data, as well as taking into account possible $\alpha$-enhancement effects in the metal-poor regime, since results will otherwise remain inconclusive or incorrect, especially if systematics cannot be fully understood or quantified. We conclude that the best available data consistently supports a non-zero, negative (i.e. metal-rich Cepheids are brighter at fixed period), metallicity dependence. \\

\section*{Acknowledgements}

We are grateful to Lennart Lindegren for providing the quasar parallax data to produce Figure 5 and for his insightful comments on the paper. LB would like to thank Paule Sonnentrucker, Chris Evans, Boris Trahin, Laura Herold, and Yukei Murakami for their continuous support. This project has received funding from the European Research Council (ERC) under the European Union's Horizon 2020 research and innovation programme (Grant Agreement No. 947660). RIA, MCR and SK are funded by the Swiss National Science Foundation through an Eccellenza Professorial Fellowship (award PCEFP2\_194638). GDS acknowledges funding from the INAF-ASTROFIT fellowship, from Gaia DPAC through INAF and ASI (PI: M.G. Lattanzi), and from INFN (Naples Section) through the QGSKY and Moonlight2 initiatives. AB thanks funding from the Anusandhan National Research Foundation (ANRF) under the Prime Minister Early Career Research Grant scheme (ANRF/ECRG/2024/000675/PMS). This research has received funding from the European Research Council under the the Horizon 2020 research and innovation program (project UniverScale, grant agreement 951549). This work has been supported by the Polish-French Marie Sk{\l}odowska-Curie and Pierre Curie Science Prize awarded by the Foundation for Polish Science. The authors acknowledge the support of the French Agence Nationale de la Recherche (ANR), under grant ANR-23-CE31-0009-01 (Unlock-pfactor). VR, MM, and ET acknowledge funding from INAF GO-GTO grant 2023 ``C-MetaLL - Cepheid metallicity in the Leavitt law'' (P.I. V. Ripepi). This research was supported by the International Space Science Institute (ISSI) in Bern/Beijing through ISSI/ISSI-BJ International Team project ID \#24-603 - ``EXPANDING Universe'' (EXploiting Precision AstroNomical Distance INdicators in the Gaia Universe), and by the International Space Science Institute (ISSI) in Bern, through ISSI International Team Project \#490, ``SH0T: The Stellar Path to the H$_0$ Tension in the Gaia, TESS, LSST and JWST Era" (PI: G. Clementini).   \\

\bibliography{Breuval_2022.bib}{}
\bibliographystyle{aasjournal}

\appendix

\section{Choice of data}

\begin{table}[h!]
\centering
\caption{Data adopted in \citetalias{MF2025} compared to the best available data. \\   }
\begin{tabular}{c c c c}
\hline
\hline
& & \citetalias{MF2025} & Best Available \\
\hline
Pulsation period  & \citetalias{MF2025} & \citet{Madore1975} & \citet{CruzReyes2023}     \\ 
for V367 Sct      & Table 1             & $\log P = 0.721$   & $\log P = 0.799$ \\
\hline
Number of SMC Cepheids & \citetalias{MF2025} & \citet{Storm2011b} & \citet{Gieren2018}  \\
with BW distances & Sect.~4.1.2 & N = 5 & N = 31 \\
\hline
Number of DEBs for & \citetalias{MF2025} & \citet{Graczyk2014} &  \citet{Graczyk2020}  \\
SMC geometric distance & Sect.~8 & N = 5 & N = 15 \\
\hline
Number of SNIa host galaxies for & \citetalias{MF2025} & \citet{Riess2016} & \citet{Riess2022a} \\
residuals of extinction curve fits & Fig.~22, 23 & N = 19 & N = 37 \\
\hline
Number of Cepheids & \citetalias{MF2025}    & \citet{Marconi2017} & \citet{Breuval2024} \\
  in SMC           & Sect.~6 & N = 9 (ground)      & N = 87 (HST) \\
\hline
Number of Cepheids & \citetalias{MF2025}    & \citet{Kochanek1997} & \citet{Li2021} \\
  in M31           & Sect.~6 & N = 30 (ground)      & N = 55 (HST) \\
\hline
Number of Cepheids & \citetalias{MF2025} & \citet{Freedman1991} & \citet{Breuval2023} \\
in M33             & Sect.~6 & N = 10 (ground) & N = 154 (HST)  \\
\hline
Number of Cepheids & \citetalias{MF2025}    & \citet{Piotto1994} & \citet{Dolphin2003}  \\
in Sextans A       & Sect.~6 &  N = 7 (ground)    & N = 82 (HST)    \\
\hline
Number of Cepheids & \citetalias{MF2025}    & \citet{Stetson1998} & \citet{Riess2022a} \\
in M101            & Sect.~6 & N = 61              & N = 260 \\
\hline
Number of Cepheids & \citetalias{MF2025}    & \citet{Riess2016} & \citet{Yuan2022} \\
in NGC 4258        & Sect.~6 & N = 141           & N = 669 \\
\hline
~ \\
\end{tabular}
\label{table:choice_data}
\end{table}

\end{document}